\documentclass[fleqn,12pt]{article}
\def\N{\rm l\!N\,}

\begin{document}
\date{}
\title{The relation between effective action and vacuum energy in 
a kappa-deformed theory}
%
\author 
{$^1$M. V. Cougo-Pinto{\footnote{\it marcus@if.ufrj.br}}, $^1$C. 
Farina{\footnote{\it farina@if.ufrj.br}} 
and $^{1,2}$J. F. M. Mendes{\footnote{\it jayme@if.ufrj.br}}
\\$^1${\it Instituto de F\' \i sica-UFRJ, CP 68528, Rio de Janeiro, RJ, 
21.941-972}\\$^2${\it 
IPD-CTEx, Av. das Am\'ericas 28.705, Rio de Janeiro, RJ, 23.020-470}\\}
\maketitle

\begin{abstract}
In a quantum field with spacetime invariance governed by the Poincar\'e
algebra the one-loop effective action is equal to the sum of
zero modes frequencies, which is the vacuum energy of the field.
The first Casimir invariant of the Poincar\'e algebra provides 
the proper time Hamiltonian in Schwinger's proper time representation of
the effective action.
We consider here a massive neutral scalar field with spacetime
invariance governed by the so called kappa-deformed Poincar\'e
algebra.  We show here that if in the kappa-deformed theory 
the first Casimir invariant of the algebra is also used as
the proper-time hamiltonian the effective action appears with a real 
and an imaginary part. The real 
part is equal to half the sum of kappa-deformed 
zero mode frequencies, which gives the vacuum energy of the kappa-deformed field. In the limit in which the deformation disappears
this real part reduces to half of the sum of zero mode frequencies
of the usual scalar field. The imaginary part is proportional to the sum
of the squares of the kappa-deformed zero mode frequencies. This
part is a creation rate of field excitations in the situations in which
it gives rise to a finite physically meaningful quantity. This is the case 
when the field is submitted to boundary conditions and properly renormalized, 
as we show in a related paper.
\end {abstract}

%

The $\kappa$-deformed Poincar\'e algebra \cite{Lukierski91} is a quantum group 
(Hopf algebra) \cite{Chaichian96} related to de Sitter and conformal algebras.
It is a deformation of the Poincar\'e algebra, which is recovered when
the limit $\kappa\rightarrow\infty$ is taken for the positive real 
parameter $\kappa$. Here we are not interested in the precise commutation
relations which define the deformed algebra, but rather in its first
Casimir invariant, which is given by
\begin{equation}\label{camadademassacapa}
{\bf P}^2-(2\kappa)^2{\rm sinh}^2(P_o/2\kappa)=-m^2, 
\end{equation}
where $m^2$ is the value of the invariant in the chosen representation.
This invariant reduces to ${\bf P}^2-P_o^2$ in the limit 
$\kappa\rightarrow\infty$, as it should be expected.

A $\kappa$-deformed field is defined in a spacetime whose symmetries are 
governed by the $\kappa$-deformed Poincar\'e algebra and accordingly 
reduces to
a relativistic quantum field in the limit $\kappa\rightarrow\infty$.
Since the continuous parameter $\kappa$ can be taken as large as we
whish the deformed algebra is well suited for the investigation of
possible small violations of relativistic symmetries. A 
$\kappa$-deformed quantum field exhibit non conservation of four momentum 
at interactions vertices and the phenomenom of creation of 
field excitations when submitted to boundary conditions. Those two
distinct phenomena point to non conservation of energy, a possibility
which should be faced once relativistic spacetime symmetries ate broken
by the deformation.

In a quantum field with spacetime invariance governed by the Poincar\'e
algebra the one-loop effective action is determined by the first 
Casimir invariant of the algebra through Schwinger's proper time 
representation of the effective action. We are interested here in the
case of a scalar field, in which Schwinger's representation is given
by
\begin{eqnarray}\label{W}
{\cal W}=-\frac{i}{2}\int_{s_o}^{\infty}\frac{ds}{s}{\rm Tr}e^{-isH},  
\end{eqnarray}
where $H={\bf P}^2 -P_o^2+m^2$ and is obtained
from the first Casimir invariant of the Poincar\'e algebra, namely:
${\bf P}^2 -P_o^2$. The effective action ${\cal W}$ is also given by 
the sum of zero modes frequencies. This sum gives
the Casimir energy of the field when it is submitted to static boundary 
conditions. This sum can also give rise to an imaginary part describing the
probability of creation of field excitations, as it occurs, {\it e.g.},
in the case of moving boundary conditions or charged vacuum submitted to
an external electric field. The method of calculating effective actions as sum of zero modes is a old one 
and the expression of the 
effective action as a sum of zero modes can actually be obtained
directly from Schwinger's proper-time representation (\ref{W})
\cite{AlbuquerqueFarina95}. 

The expression (\ref{W}) for a $\kappa$-deformed scalar field requires
\begin{equation}\label{Hcapa}
H={\bf P}^2 -(2\kappa)^2{\rm sinh}^2(P_o/2\kappa)+m^2, 
\end{equation}
in accordance with the first Casimir invariant (\ref{camadademassacapa})
of the $\kappa$-deformed Poincar\'e algebra. For the field submitted
to static boundary conditions describing confinement between parallel plates
this effective action has a real and imaginary part \cite{{Schwinger51}, {GribMamayevMostepanenko94}, {GreinerMullerRafelski85}}. 

The real part is a Casimir energy of the $\kappa$-deformed field and actually reduces to
the Casimir energy of the usual scalar field in the limit 
$\kappa\rightarrow\infty$. The imaginary part gives a probability of 
creation of field excitations which disappears in the limit
$\kappa\rightarrow\infty$ or in the limit when the boundary conditions
are turned off (infinite separation between the plates). This the
above mentioned creation mechanism steaming from the $\kappa$-deformation
and totally absent in the usual field theories. In order to further
clarify these results we address here the relation between the
effective action (\ref{W}) and the sum of zero modes in the case of
a $\kappa$-deformed scalar field. The scalar field is submitted to 
boundary conditions in view of future applications of the results
developed here. These results can be obtained in the absence of 
boundary conditions without further efforts. 

Let us consider the $\kappa$-deformed scalar field submitted to
Dirichlet boundary conditions on two large parallel plates of side
$\ell$ and separation $a$ ($a\ll\ell$). We take the ${\cal OZ}$-axis
perpendicular to the plates and the eigenvalues of $P_z$ in 
(\ref{Hcapa}) are accordingly given by $n\pi/a$ ($n\in\N$).
We start by taking (\ref{Hcapa}) into the effective action
(\ref{W}) with regularization parameters $\epsilon$ and $\nu$ to
obtain: 
\begin{eqnarray}\label{acao-efetiva}
{\cal W}=-\frac{i}{2}\int_{0}^{\infty}\frac{ds}{s}s^\nu
{\rm Tr}\exp\{-is[{\bf P}^2 -(2\kappa)^2{\rm sinh}^2(P_o/2\kappa)+m^2-i\epsilon]\}
\; ,
\end{eqnarray}
where $\epsilon$ is a positive infinitesimal and $\nu$ is large enough to
render the integral well defined. By taking the boundary conditions in
consideration the above expression acquires the form:
\[
{\cal W}=-\frac{i}{2}\frac{T\ell^2}{(2\pi)^3}\int_{0}^{\infty}\frac{ds}{s}s^\nu
\sum_{n=1}^\infty\int{d\omega}\int{dp_1}\int{dp_2}\times
\]
\begin{equation}
\times\exp\{-is[p_{1}^{2}+p_{2}^{2}
+(n\pi/a)^2-(2\kappa)^2{\rm sinh}^2(\omega/2\kappa)+m^2-i\epsilon]\}
\end{equation}

By taking the derivative of this expression in relation to $m^2$, 
\[
\frac{1}{T\ell^2}\frac{\partial{\cal W}}{\partial{m^2}}=-\frac{1}{2}\frac{1}{(2\pi)^3}
\sum_{n=1}^\infty\int{d\omega}\int{dp_1}\int{dp_2}\int_{0}^{\infty}dss^{\nu}\times
\]
\begin{equation}
\times\exp\{-is[p_{1}^{2}+p_{2}^{2}+(n\pi/a)^2-(2\kappa)^2{\rm sinh}^2(\omega/2\kappa)+m^2
-i\epsilon]\}, 
\end{equation}
and using Euler's Gamma function integral representation \cite{GradshteynRyzhik65} 
%
%
we arrive at: 
\begin{eqnarray}\label{derondWderondm2dois}
\frac{1}{T\ell^2}\frac{\partial{\cal W}}{\partial{m^2}}=-\frac{1}{2}i^{\nu +1}
\Gamma({\nu+1})(2\kappa^2)^{-(\nu+1)}\sum_{n=1}^\infty\int\int\frac{dp_1dp_2}{(2\pi)^2}
\int\frac{d\omega}{2\pi}\left[{\rm cosh}\left(\frac{\omega}{\kappa}\right)
-\beta\right]^{-(\nu+1)}\label{energia-reg.1}, 
\end{eqnarray}
where
\begin{eqnarray}\label{definicaodebeta}
\beta=1+\frac{1}{2\kappa^2}[p_{1}^{2}+p_{2}^{2}+(n\pi/a)^2+m^2-i\epsilon]
\; .
\end{eqnarray}
We retain for a while the regulator $\epsilon$ to perform the integration over $\omega$. We consider $\nu$ to be a positive integer to obtain:
\begin{eqnarray}\label{derondWderondm2tres}
\frac{1}{T\ell^2}\frac{\partial{\cal W}}{\partial{m^2}}&=&\frac{i^{\nu+1}}{2}
\Gamma({\nu+1})(2\kappa^2)^{-(\nu+1)}\times\nonumber\\
&&\times\sum_{n=1}^\infty
\int\int\frac{dp_1dp_2}{(2\pi)^2}\frac{1}{2\pi}
\frac{\kappa}{\nu!}
\frac{\partial^\nu}{\partial\beta^\nu}
\left[\frac{2\pi{i}}{\sqrt{\beta^2-1}}-\frac{2}{\sqrt{\beta^2-1}}
\log{(\beta+\sqrt{\beta^2-1})}\right]\; ,
\end{eqnarray}
where we have finally taken $\epsilon\rightarrow 0$. From the definition
(\ref{definicaodebeta}) we have that  
$\partial{\cal W}/\partial\beta=2\kappa^2\partial{\cal W}/\partial{m^2}$
This identity can be used to eliminate the derivative in relation to the square mass in (\ref{derondWderondm2tres}), which can be integrated in $\beta$
to become:
\begin{eqnarray}\label{Wfinal}
\frac{1}{i^\nu}\frac{\cal W}{T\ell^2}=-\frac{\kappa}{(2\kappa^2)^{\nu}}
\sum_{n=1}^\infty\int\int\frac{dp_1dp_2}{(2\pi)^3}\int_{\infty}^{\beta}{d\beta}
\frac{\partial^\nu}{\partial\beta^\nu}
\left[\frac{\pi}{\sqrt{\beta^2-1}}+\frac{i}{\sqrt{\beta^2-1}}
\log{(\beta+\sqrt{\beta^2-1})}\right]. 
\end{eqnarray}
Now we understand that the $\nu$ derivatives in relation to $\beta$
has been taken in the integrand of (\ref{Wfinal}). The resulting 
expression is a function of $\nu$ that we submit to an analytical
continuation. In this way the limit $\nu\rightarrow 0$ can be taken
after the subtraction of spurious terms in order to arrive at the 
physical quantities. The identification of spurious terms depends
on the specific problem in consideration. Here we will be content
in showing that (\ref{Wfinal}) is properly regularized by the 
parameter $\nu$ and that the elimination of the regularization
gives us the relation between the effective action 
(\ref{W}) and the sum of zero modes, both without regularization.
This relation is obtained by taking the value $\nu=0$ in (\ref{Wfinal})
to arrive at:
\begin{equation}\label{W=omega+iomega2}
-\frac{\cal W}{T\ell^2}=\sum_{n=1}^\infty
\int\int\frac{dp_1dp_2}{(2\pi)^2} \frac{1}{2}\omega(p_1,p_2,n)+
\frac{i}{\pi\kappa}\sum_{n=1}^\infty
\int\int\frac{dp_1dp_2}{(2\pi)^2}\frac{1}{4}\omega^2(p_1,p_2,n)
\end{equation}
where $\omega$ is the frequency given by the mass-shell condition
derived from (\ref{camadademassacapa}):
\begin{equation}\label{relacaodedispersaocapa}
\omega(p_1,p_2,n)=2\kappa \sinh^{-1}\left[
\frac{1}{2\kappa}\sqrt{p_{1}^{2}+p_{2}^{2}+(\pi n/a)^2+m^2}\right] \; .
\end{equation}
If we prefer we can write (\ref{W=omega+iomega2}) using box normalization
in order to discretize all the components of momentum:
\begin{equation}\label{W=omega+iomega2discreto}
-\frac{\cal W}{T}=\sum_{\bf p}\frac{1}{2}\omega_{\bf p}
+\frac{i}{\pi\kappa}
\sum_{\bf p}\frac{1}{4}\omega_{\bf p}^2. 
\end{equation}
The expression (\ref{W=omega+iomega2}) is our main result, which answers 
the question 
of what is the relation  between the effective action (\ref{W}) and the sum 
of zero 
modes in the case of a $\kappa$-deformed scalar field. The result 
(\ref{W=omega+iomega2}) shows that the effective
action ${\cal W}$ as given in (\ref{W}) has a real part given by the sum
of half frequencies, as in the non-deformed case, although the frequencies
to be summed in the present case are given by the $\kappa$-deformed
expression (\ref{relacaodedispersaocapa}). Contrary to the non-deformed
case the effective action has also an imaginary part, which is given
by a sum of squared half frequencies divided by the deformation parameter
$\kappa$. In the limit $\kappa\rightarrow\infty$ in which the deformation
disappears the imaginary part goes to zero and the real part goes to 
the sum of half non-deformed frequencies:  
\begin{equation}
\lim_{\kappa\rightarrow\infty}\frac{\cal W}{T}=
-\sum_{\bf p}\frac{1}{2}\sqrt{{\bf p}^2+m^2} \; ,
\end{equation}
which is exactly what should be expected. The imaginary part in
(\ref{W=omega+iomega2}) is responsible for the creation of field
excitations when the $\kappa$-deformed field is submitted to 
boundary conditions. This part is in total agreements with 
previously obtained results \cite{CougoPinto-Farina97} and will 
be further investigated in a companion paper.

We are left now with the task of proving that the non-regularized
result (\ref{W=omega+iomega2discreto}) is on firm ground. To this 
purpose we show that (\ref{Wfinal}) is properly regularized by the 
parameter $\nu$. This is a rather technical manipulation of inequalities 
and estimation of integrals that we present bellow, in a mode of appendix. 
We want to show that (\ref{Wfinal}) is properly regularized, {\it i.e.}, that
there exists $\nu$ such that the integral in (\ref{Wfinal}) are well defined.
%
%
Concerning the first term in the integrand of (\ref{Wfinal}) we want to show
that:
\begin{equation}
\left|\frac{\partial^\nu}{\partial\beta^\nu}\frac{1}{\sqrt{\beta^2-1}}\right|{\leq}\left|\frac{\partial^\nu}{\partial\beta^\nu}\frac{1}{\beta-\beta^{-1}}\right|.
\label{leq-beta} 
\end{equation}
By using the series expansions: 
\begin{eqnarray}\label{desig.1}
\frac{\partial^\nu}{\partial\beta^\nu}\frac{1}{\sqrt{\beta^2-1}}&=&(-1)^\nu
\sum_{j=0}^{\infty}\frac{(2j+\nu)!}{(2^jj!)^2}\beta^{-(2j+\nu+1)},\\
\frac{\partial^\nu}{\partial\beta^\nu}\frac{1}{\beta-\beta^{-1}}&=&
(-1)^\nu\sum_{j=0}^{\infty}\frac{(2j+\nu)!}{(2j)!}\beta^{-(2j+\nu+1)},
\end{eqnarray}
we reduce the verification of inequality (\ref{leq-beta}) to the verification
of the following simpler inequality:
\begin{equation}\label{2j<2j}
\frac{1}{(2^jj!)^2}\leq\frac{1}{(2j)!}\; .
\end{equation}
Since 
\begin{eqnarray}
2^{2j}=\sum_{n=0}^{2j}{2j\atopwithdelims()n}=
{2j\atopwithdelims()j}+\sum_{n=0\;(n\neq{j})}^{2j}{2j\atopwithdelims()n}, 
\end{eqnarray}
and all the quantities in the above equation are positive, we conclude that 
\begin{eqnarray}
{2j\atopwithdelims()j}\,{\leq}\,\,2^{2j}\; ,
\end{eqnarray}
which shows the correctness of (\ref{2j<2j}) and thus the validity of
the inequality (\ref{leq-beta}). We now use this inequality in the real
part of (\ref{Wfinal}) to obtain: 
\begin{eqnarray}
\left|\Re\left\{-\frac{1}{i^\nu}\frac{\cal W}{T\ell^2}\right\}\right|{\leq}\left|
\frac{\kappa\pi}{(2\kappa^2)^{\nu}}\sum_{n=1}^\infty\int\int\frac{dp_1dp_2}{(2\pi)^3}\int_{\infty}^{\beta}{d\beta}\left|
\frac{\partial^\nu}{\partial\beta^\nu}\frac{1}{\beta-\beta^{-1}}\right|\right|. \label{mod.real/E}
\end{eqnarray}
%
We have now to calculate the $\nu$'nd derivative in (\ref{mod.real/E}), 
which result is given below:    
\begin{eqnarray}
\frac{\partial^\nu}{\partial\beta^\nu}\frac{1}{\beta-\beta^{-1}}=\frac{(-1)^\nu{\nu !}}{2}
\left[\frac{1}{(\beta+1)^{\nu+1}}+\frac{1}{(\beta-1)^{\nu+1}}\right]\label{dif.f}. 
\end{eqnarray}
Substituting this result in (\ref{mod.real/E}) 
we obtain:
\begin{eqnarray}
\left|\Re\left\{-\frac{1}{i^\nu}\frac{\cal W}{T\ell^2}\right\}\right|\leq
\left|\frac{\kappa\pi{\nu !}}{2{(2\kappa^2)}^\nu}\sum_{n=1}^\infty\int\int\frac{dp_1dp_2}{(2\pi)^3}
\int_{\infty}^{\beta}{d\beta}\left[\frac{1}{(\beta+1)^{\nu+1}}+\frac{1}{(\beta-1)^{\nu+1}}\right]
\right|. 
\end{eqnarray}
%
We see in this expression that the integrals $p_1$ and $p_2$ are well defined
for $\nu\geq{1}$. The integration over $\beta$ reduces the expression to:  
\begin{equation}
\left|\Re\left\{-\frac{1}{i^\nu}\frac{\cal W}{T\ell^2}\right\}\right|{\leq}\left|-
\frac{\kappa\pi{(\nu-1)!}}{2(2\kappa^2)^\nu}\sum_{n=1}^\infty\int\int
\frac{dp_1dp_2}{(2\pi)^3}\left[\frac{1}{(\beta+1)^\nu}+\frac{1}{(\beta-1)^\nu}\right]
\right|, 
\end{equation}
or, 
\begin{equation}
\left|\Re\left\{-\frac{1}{i^\nu}\frac{\cal W}{T\ell^2}\right\}\right|{\leq}\left|-
\frac{\kappa\pi{(\nu-1)!}}{(2\kappa^2)^\nu}\sum_{n=1}^\infty\int\int\frac{dp_1dp_2}{(2\pi)^3}
\frac{1}{(\beta-1)^\nu}
\right|. \label{desig.energia} 
\end{equation}
From the definition (\ref{definicaodebeta}) of $\beta$ we have 
\[
\beta=
\frac{p_{\parallel}^{2}}{2\kappa^2}+
\alpha_{n}(a,m,{\kappa}), 
\]
where: 
\[
p_{\parallel}^{2}=p_{1}^{2}+p_{2}^{2}\; , 
\]
and
\[
\alpha_{n}(a,m,{\kappa})=1+\frac{1}{2\kappa^2}\left(m^2
+\left(\frac{n\pi}{a}\right)^2\right)\; ,
\] 
%
%
being clear that:
\[
\alpha_{n}(a,m,{\kappa})\geq{1} 
\]
We then write (\ref{desig.energia})as  
\begin{equation}
\left|\Re\left\{-\frac{1}{i^\nu}\frac{\cal W}{T\ell^2}\right\}\right|{\leq}\left|-
\frac{\kappa\pi{(\nu-1)!}}{(2\kappa^2)^\nu}\sum_{n=1}^\infty\int_{0}^{2\pi}\frac{d\theta}{({2\pi})^3}
\int_{0}^{\infty}\frac{dp_{\parallel}p_{\parallel}}{\left[\frac{p_{\parallel}^{2}}{2\kappa^2}
+\alpha_{n}(a,m,{\kappa})\right]^\nu}
\right|\label{desig.energia2}.  
\end{equation}
If ${p_{\parallel}^{2}}/{2\kappa^2}+\alpha_{n}(a,m,{\kappa})=x$, with $\nu>1$ 
(to avoid divergences), we have: 
\begin{equation}\label{intP||}
\int_{0}^{\infty}\frac{dp_{\parallel}p_{\parallel}}{\left[\frac{p_{\parallel}^{2}}
{2\kappa^2}+\alpha_{n}(a,m,{\kappa})\right]^\nu}=
=-\frac{1}{\nu-1}\frac{1}{x^{\nu-1}}\Bigg|_{\alpha_n}^{\infty}=
\frac{1}{\nu-1}\frac{1}{[\alpha_{n}(a,m,{\kappa})]^{\nu-1}}
\end{equation}
By assuming that $\nu>1$ we can perform the integration on $p_{\parallel}$ to obtain:
\begin{equation}
\left|\Re\left\{-\frac{1}{i^\nu}\frac{\cal W}{T\ell^2}\right\}\right|{\leq}\left|-
\frac{\kappa(\nu-2)!}{4\pi(2\kappa^2)^\nu}\sum_{n=1}^\infty\frac{1}{[\alpha_{n}(a,m,{\kappa})]^{\nu-1}}
\right|\label{desig.energia3},  
\end{equation}
or, 
\begin{equation}
\left|\Re\left\{-\frac{1}{i^\nu}\frac{\cal W}{T\ell^2}\right\}\right|{\leq}\left|-
\frac{\kappa(\nu-2)!}{4\pi(2\kappa^2)^\nu}\sum_{n=1}^\infty\frac{1}{\left[1+\frac{1}{2\kappa^2}
\left(m^2+\left(\frac{n\pi}{a}\right)^2\right)\right]^{\nu-1}}
\right|.\label{desig.energia4} 
\end{equation}
Right side of (\ref{desig.energia4}) has a finite value and is a representation of the Epstein function. This function has an analytical continuation \cite{AmbjornWolfram81} and so we can choose any value for the variable $\nu$. 
If $2(\nu-1)\!\!>\!1$, 
or $\nu\!>\!3/2$, we conclude that there exists $\nu$ such that real part of the Schwinger's 
effective action is regularized by the power $s^\nu$ in (\ref{acao-efetiva}).



Now we have to verify if the imaginary part of the effective action, equation (\ref{Wfinal}),
is already regularized. 
\begin{equation}
\Im\left\{-1\frac{1}{i^\nu}\frac{\cal W}{T\ell^2}\right\}=\frac{\kappa}{(2\kappa^2)^{\nu}}
\sum_{n=1}^\infty\int\int\frac{dp_1dp_2}{(2\pi)^3}\int_{\infty}^{\beta}{d\beta}\frac{\partial^\nu}{\partial\beta^\nu}
\frac{\log{(\beta+\sqrt{\beta^2-1})}}{\sqrt{\beta^2-1}}\label{energia-imaginaria}.
\end{equation}
As made for the real part, we have to maximize the imaginary part in 
(\ref{energia-imaginaria}). To make that, we try to find a function 
$M\!:\beta\rightarrow{M}(\beta)$ that satisfy the inequality given by: 
\[
\int_{\infty}^{\beta}{d\beta}\frac{\partial^\nu}{\partial\beta^\nu}
\frac{\log{(\beta+\sqrt{\beta^2-1})}}{\sqrt{\beta^2-1}}\leq\int_{\infty}^{\beta}{d\beta}\left|
\frac{\partial^\nu}{\partial\beta^\nu}
\frac{\log{(\beta+\sqrt{\beta^2-1})}}{\sqrt{\beta^2-1}}\right|\leq\int_{\infty}^{\beta}{d\beta}
\left|M(\beta)\right|
\] 
Defining $g(\beta)=1/\sqrt{\beta^2-1}$, and $h(\beta)=\log(\beta+\sqrt{\beta^2-1})$, 
where, $\partial{h(\beta)}/\partial\beta=g(\beta)$, 
the derivative of order $\nu$ of a product of two functions can be calculated from: 
\[
\frac{\partial^\nu}{\partial\beta^\nu}(g(\beta)h(\beta))=\sum_{l=0}^{\nu}{\nu\atopwithdelims()l}
\frac{\partial^{\nu-l}g(\beta)}{\partial\beta^{\nu-l}}\frac{\partial^{l}h(\beta)}{\partial\beta^{l}}=
\]
\begin{equation}
=\frac{\partial^{\nu}g(\beta)}{\partial\beta^{\nu}}h(\beta)+\sum_{l=1}^{\nu}{\nu\atopwithdelims()l}
\frac{\partial^{\nu-l}g(\beta)}{\partial\beta^{\nu-l}}
\frac{\partial^{l}h(\beta)}{\partial\beta^{l}}\label{M-dif}
\end{equation}
But  
\[
\frac{\partial^{l}h(\beta)}{\partial\beta^{l}}=\frac{\partial^{l-1}}{\partial\beta^{l-1}}
\frac{\partial{h(\beta)}}{\partial\beta}=\frac{\partial^{l-1}g(\beta)}{\partial\beta^{l-1}}
\]
Then, we can rewrite the (\ref{M-dif}) as: 
\[
\frac{\partial^\nu}{\partial\beta^\nu}(g(\beta)h(\beta))=
\frac{\partial^{\nu}g(\beta)}{\partial\beta^{\nu}}h(\beta)+\sum_{l=1}^{\nu}{\nu\atopwithdelims()l}
\frac{\partial^{\nu-l}g(\beta)}{\partial\beta^{\nu-l}}
\frac{\partial^{l-1}g(\beta)}{\partial\beta^{l-1}}, 
\]
and also knowing that, 
\[
\left|\frac{\partial^\nu}{\partial\beta^\nu}(g(\beta)h(\beta))\right|\leq\left|
\frac{\partial^{\nu}g(\beta)}{\partial\beta^{\nu}}h(\beta)\right|+\left|
\sum_{l=1}^{\nu}{\nu\atopwithdelims()l}\frac{\partial^{\nu-l}g(\beta)}{\partial\beta^{\nu-l}}
\frac{\partial^{l-1}g(\beta)}{\partial\beta^{l-1}}\right|\leq\nonumber
\]
\begin{equation}
\leq\left|\frac{\partial^{\nu}g(\beta)}{\partial\beta^{\nu}}h(\beta)\right|+
\sum_{l=1}^{\nu}{\nu\atopwithdelims()l}\left|
\frac{\partial^{\nu-l}g(\beta)}{\partial\beta^{\nu-l}}\right|\left|
\frac{\partial^{l-1}g(\beta)}{\partial\beta^{l-1}}\right|\label{desig.gh}, 
\end{equation}
we are ready to search for functions $g$ and $h$ that maximize the imaginary part 
of the Casimir energy. The function $g$ and its derivatives of the $\nu$ order were 
already maximized by the function $f\!:{\beta}\rightarrow{f}(\beta)=
(\beta-\beta^{-1})^{-1}$. So, let's analyze now the case of $h$, verifying if exists 
$\gamma$ such that:  
\[
0{\leq}h(\beta)=\log({\beta+\sqrt{\beta^2-1}})\leq\log({\gamma\beta}), 
\]
In other words, 
\[
\gamma\geq1+\frac{\sqrt{\beta^2-1}}{\beta}. 
\]
in order to simplify and choose $\gamma$ as being a numerical constant, we have to take 
the maximum of $1+\sqrt{\beta^2-1}/{\beta}$ that is 2. So we conclude that, 
\[
0\leq{h(\beta)}=\log({\beta+\sqrt{\beta^2-1}})<\log({2\beta})<\log(e^{2\beta})=2\beta\;\;
\rightarrow 
\]
\[
\rightarrow\;\;h(\beta)=\log({\beta+\sqrt{\beta^2-1}})<2\beta. 
\]
Then, $\exists\;\;\sigma\;\supset\!\!\!-\; \sigma\beta\geq\log(\beta+\sqrt{\beta^2-1})$. 
We can now rewrite the inequality (\ref{desig.gh}) as, 
\[
\left|\frac{\partial^\nu}{\partial\beta^\nu}
\frac{\log({\beta+\sqrt{\beta^2-1)}}}{\sqrt{\beta^2-1}}
\right|\leq
\]
\begin{equation}
\leq\left|\sigma\beta\frac{\partial^{\nu}}{\partial\beta^{\nu}}\frac{1}{\beta-\beta^{-1}}\right|
+\sum_{l=1}^{\nu}{\nu\atopwithdelims()l}\left|
\frac{\partial^{\nu-l}}{\partial\beta^{\nu-l}}\frac{1}{\beta-\beta^{-1}}
\right|\left|\frac{\partial^{l-1}}{\partial\beta^{l-1}}\frac{1}{\beta-\beta^{-1}}\right|
=\left|M(\beta)\right|, 
\end{equation}
using equation (\ref{dif.f}) we may write,

\[
\left|\frac{\partial^\nu}{\partial\beta^\nu}
\frac{\log({\beta+\sqrt{\beta^2-1)}}}{\sqrt{\beta^2-1}}
\right|\leq\sigma\beta\frac{\nu!}{2}\left[\frac{1}{(\beta+1)^{\nu+1}}+\frac{1}{(\beta-1)^{\nu+1}}\right]
+
\]
\[
+\sum_{l=1}^{\nu}{\nu\atopwithdelims()l}\frac{(l-1)!}{2}\left[\frac{1}{(\beta+1)^{l}}
+\frac{1}{(\beta-1)^l}\right]\frac{(\nu-l)!}{2}
\left[\frac{1}{(\beta+1)^{\nu-l+1}}+\frac{1}{(\beta-1)^{\nu-l+1}}\right]=
\]
\[
=\sigma\beta\frac{\nu!}{2}\left[\frac{1}{(\beta+1)^{\nu+1}}+\frac{1}{(\beta-1)^{\nu+1}}\right]+
\]
\[
+\sum_{l=1}^{\nu}\frac{\nu!}{4l}\left[\frac{1}{(\beta+1)^{\nu+1}}+\frac{1}{(\beta-1)^{\nu+1}}
+\frac{1}{(\beta+1)^{\nu+1}}\left(\frac{\beta+1}{\beta-1}\right)^l+
\frac{1}{(\beta-1)^{\nu+1}}\left(\frac{\beta-1}{\beta+1}\right)^l\right]<
\]
\[
<\sigma\beta\frac{\nu!}{(\beta-1)^{\nu+1}}+
\sum_{l=1}^{\nu}\frac{\nu!}{(\beta-1)^{\nu+1}}=\frac{{\nu}!(\nu+\sigma\beta)}{(\beta-1)^{\nu+1}}
\]
In this sense, we can assert that, 
\begin{equation}
\Im\left\{-\frac{1}{i^\nu}\frac{\cal W}{T\ell^2}\right\}<\left|\frac{\kappa{\nu!}}{(2\kappa^2)^\nu}
\sum_{n=1}^\infty\int\int\frac{dp_1dp_2}{(2\pi)^3}\int_{\infty}^{\beta}{d\beta}\frac{(\nu+\sigma\beta)}{(\beta-1)^{\nu+1}} 
\right|,
\end{equation}
or, considering $\nu>1$,  
\begin{equation}
\Im\left\{-\frac{1}{i^\nu}\frac{\cal W}{T\ell^2}\right\}<\left|\frac{\kappa{\nu!}}{(2\kappa^2)^\nu}
\sum_{n=1}^\infty\int\int\frac{dp_1dp_2}{(2\pi)^3}\left[\frac{\sigma}{(\nu-1)(\beta-1)^{\nu-1}}
+\frac{\sigma+\nu}{\nu (\beta-1)^{\nu}}\right]\right|
\end{equation}
The integrals over $p_1$ and $p_2$ using the expression of $\beta(p_1,p_2,n,a)$ given 
in (\ref{intP||}) was already done, then: 
\[
\Im\left\{-\frac{1}{i^\nu}\frac{\cal W}{T\ell^2}\right\}<\Biggl|\frac{\kappa{\nu!}}{(2\kappa^2)^\nu}
\sum_{n=1}^\infty\frac{1}{(2\pi)^2}\Biggl\{\frac{\sigma}{(\nu-1)(\nu-2)\left\{\frac{1}{2\kappa^2}
\left[m^2+\left(\frac{n\pi}{a}\right)^2\right]\right\}^{\nu-2}}+
\]
\[
+\frac{\sigma+\nu}{\nu (\nu-1)\left\{\frac{1}{2\kappa^2}\left[m^2+
\left(\frac{n\pi}{a}\right)^2\right]\right\}^{\nu-1}}\Biggr\}\Biggr|. 
\]
Since these sums converge if $\nu>5/2$, the regularization also works to the imaginary part. 

The intersection of all the possible values for $\nu$ is given from the inequality $\nu>5/2$,
 which shows that there exists $\nu$ such that the effective action can be regularized by a 
 power regularization introduced in (\ref{acao-efetiva}). 

Despite the power regularization has been proved to be valid to calculate the Schwinger's 
effective action, if we want to calculate the Casimir energy from 
${\cal E}=-{\cal W}/T$ \cite{Schwinger51} and from the non-regularized sum of zero modes 
given by (\ref{W=omega+iomega2discreto}), it is convenient, for practical purposes, to 
use another type of regularization. A simple one given by $\exp\{-\epsilon |{\bf k}|\}$ 
would be enough.

\end{document}